BALZAN MARIO V

# ASSESSING ECOSYSTEM SERVICES FOR EVIDENCE-BASED NATURE-BASED SOLUTIONS INTERVENTIONS

ABSTRACT

The term 'nature-based solutions' has often been used to refer to adequate green infrastructure, which is cost-effective and simultaneously provides environmental, social and economic benefits, through the delivery of ecosystem services, and contributes to build resilience. This paper provides an overview of the recent work mapping and assessing ecosystem services in Malta and the implications for decision-making. Research has focused on the identification and mapping of ecosystems, and ecosystem condition, the capacity to deliver key ecosystem services and the actual use (flow) of these services by local communities leading to benefits to human well-being.

The integration of results from these different assessments demonstrates several significant synergies between ecosystem services, indicating multifunctionality in the provision of ecosystem services leading to human well-being. This is considered as key criterion in the identification of green infrastructure in the Maltese Islands. A gradient in green infrastructure cover and ecosystem services capacity is observed between rural and urban areas but ecosystem services flow per unit area was in some cases higher in urban environments. These results indicate a potential mismatch between ecosystem service demand and capacity but also provide a scientific baseline for evidence-based policy which fosters the development of green infrastructure through 'nature-based' innovation promoting more specific and novel solutions for landscape and urban planning.

## INTRODUCTION

During the last five years, Europe has seen a surge of research and policy initiatives on ecosystem services (Maes & Jacobs, 2017), the latter being defined as the direct and indirect contributions of ecosystems to human well-being (De Groot et al., 2010). Ecosystem services research connects ecology with human-wellbeing and economy by analysing the links between ecosystems and the benefits people receive from nature (Potschin & Haines-Young 2016). The contributions of ecosystems to the attainment of social and economic policy objectives has been recognised by the European Union policy. The EU Biodiversity Strategy to 2020 sets important targets for the development of knowledge about ecosystems, their services and values in the national territory of member states and the integration of these values into accounting and reporting systems at EU and national level. Building on this, the EU Green Infrastructure Strategy recognises the pivotal contributions of ecosystems for more systematic economic solutions (Maes & Jacobs 2017).

The important role played by green infrastructure for sustainable economic growth is also recognised by Malta's National Green Economy Strategy (2015). Green infrastructure, defined as a strategically planned network of natural and semi-natural areas that is multifunctional, providing multiple services and benefits whilst also protecting biodiversity in rural and urban areas. The multifunctionality of green infrastructure is also, indirectly, recognised by Malta's Strategic Plan for Environment and Development (SPED, 2015) which includes in its objectives the protection of existing recreational areas to improve social cohesion and human health, whilst supporting the strengthening of the existing ecological network and recognising the important contribution of urban green spaces. Malta's National Biodiversity Strategy and Action Plan (NPSAP) 2012-2020 also recognises the important value of ecosystems and their services and aims to develop the existing thematic knowledge base (Target 18), integrate this in national policies as well as decision-making and planning processes (Target 2), and restore at least 15% of degraded ecosystems and for the essential services provided vulnerable ecosystems to be safeguarded (Target 13).

The operationalisation of the ecosystem service concept is a key objective for researchers and practitioners working at the interface of science and policy. In its most rudimentary sense the idea is to mainstream ecosystem services into decision-making because there are a range of nature-based solutions to societal challenges (Potschin-Young et al., 2016). However, a number of challenges to the implementation and uptake of the ecosystem services concept exist, and the mapping and assessing of ecosystems and their services has been implemented with varying success across the EU member states (Kopperoinen, Varumo, & Maes, 2018). The most successful EU countries having implemented nation-wide ecosystem service assessment and developed methodologies and tools for their use in decision-making. The assessment and mapping of ecosystem services, and their integration in decision-making, presents a number of challenges which include the lack of an accepted and coherent approach and frameworks and the limited availability of empirical data (Burkhard, Maes, et al., 2018). The implementation of the ecosystem assessments in Malta (e.g. as reviewed by Mallia and Balzan, 2015), is associated with a number of challenges, which include the availability of biodiversity and ecosystem services spatial data that cover the territory at adequate resolutions (Balzan, Caruana, & Zammit, 2018). This appears to be a common situation for small island states across the globe, as demonstrated by a recent review about small island ecosystem services which indicates that studies that carried out a biophysical quantification of ecosystem services, investigated their spatial variation, arising synergies and trade-offs, or assessed the socio-cultural and economic value of island ecosystem services are rather limited. On the contrary, most of the studies dealt with the management of pressures, and externalities and trade-offs associated with the use of ecosystems and their services (Balzan, 2018; Balzan, Potschin, & Haines-Young, 2018). This is congruent with the view that sees nature as a resource to be exploited for temporary economic successes (Maes & Jacobs, 2017).

The assessment of ecosystem services requires a complete understanding of the flow of services from ecosystems to society, and consequently the use of different indicators that are based on meaningful science is required for an effective implementation of the ecosystem services concept (Potschin & Haines-Young, 2016; Villamagna, Angermeier, & Bennett, 2013). This contribution provides an overview of the recent research activities mapping and assessing ecosystem services at a national scale, and briefly discusses the implications for decision-making.

CONCEPTUAL FRAMEWORK

The ecosystem service framework adopted in this work (Figure 1), builds on existing widely-used frameworks (Potschin & Haines-Young, 2016; Villamagna et al., 2013), and distinguishes between different components along the ecosystem services delivery chain (Table 1). To understand the relationship between people and ecosystems, we need to identify both the functional characteristics of ecosystems that give rise to services and the benefits and values which arise from these (Potschin & Haines-Young, 2016). Assessments carried out distinguish between ecosystem services capacity and flow, with the capacity being defined as the potential of ecosystems to provide service, while the flow refers to the actual production and mobilisation of the service. The ecosystem service demand depends on the beneficiaries' preferences for specific ecosystem service attributes (Potschin-Young, Burkhard, Czúcz, & Santos-Martín, 2018). In a final 'societal input' feedback loop, through decision-making, societies manage drivers of change impacting indirectly on ecosystem through the action of pressures (Nassl & Löffler, 2015).

The implementation of ecosystem-based approaches and measures which use nature's multiple services as 'nature-based solutions' to societal challenges is considered as being part of the governance regime within this framework as it effectively impacts on the drivers and pressures acting on ecosystems whilst promoting the provision of ecosystem services and improving resilience. Nature-based solutions operationalise the concept of the ecosystem services in real-world situations (Faivre, Fritz, Freitas, de Boissezon, & Vandewoestijne, 2017). Nature-based solutions are considered as being part of the green infrastructure since they use biodiversity and ecosystem services as part of an overall adaptation strategy.

MAPPING AND ASSESSING ECOYSTEM SERVICES

Recent EU-funded research has permitted the development of flexible ecosystem services mapping and assessment methodologies, and their testing in representative thematic and biome-oriented case-studies. This is expected to promote local ecosystem service assessments required for spatial planning, agriculture, climate, water and biodiversity policy (Burkhard, Maes, et al., 2018). The EU-funded Horizon 2020 project ESMERALDA[i] (Enhancing ecosystem services mapping for policy and decision making) has developed a 7-step mapping and assessment of ecosystem services (MAES) implementation plan that:

1.  identifies questions that stakeholders have and which may be answered by the MAES;
2.  identifies relevant stakeholders, for instance from science, policy or society, that are in a position to deal with these question;
3.  establishes a network involving the relevant stakeholders;
4.  implements the mapping and assessment process according to the available knowledge and data aspects;
5.  implements case-studies to test the MAES approaches;
6.  develops a user-oriented dissemination and communication of ecosystem service mapping and assessment outcomes is implemented;

7. integrates ecosystem services in decision-making in policy, business and practice (Burkhard, Sapundzhieva, et al., 2018).

A case-study has been carried out to map and assess ecosystem services in the Maltese Islands and to test assessment methodologies (Balzan, Caruana, et al., 2018). This study consists of a first assessment of the capacity and flow of ecosystem services in the Maltese Islands, and has been carried out in order to analyse the spatial variation of ecosystem services, identify service hotspots and explore the impact of policies and developments on the ecosystems' capacity to provide key ecosystem services. The assessment builds on an analysis of the status of active research and policy initiatives, prerequisites and needs at start of the project. This review identified the ecosystems covered in the country, listed some of most relevant ecosystem services, and identified stakeholders that may be involved in the mapping and assessment process (Mallia & Balzan, 2015).

The mapping and assessment of ecosystem services is largely dependent on the availability of land cover land use (LULC) datasets at adequate resolutions and of data about the condition of ecosystems that is based on information about drivers, pressures and the impacts on structure and function of ecosystems. Balzan et al. (2018a) identified spatial data availability about ecosystems and their condition as being an important limitation for the mapping and assessment of ecosystem services. Similar observations have been made elsewhere suggesting that there is often a lack of local-scale data for the implementation of ecosystem service assessments for decision-making (Burkhard, Kroll, Nedkov, & Müller, 2012).

During the implemented case-study, a LULC map was created through the use of Sentinel 2 satellite images provided by Copernicus (Drusch et al., 2012). Following the adoption of the conceptual framework (Figure 1), a number of indicators that may be used to assess key ecosystem services in the Maltese Islands were identified (Table 2). An ecosystem service indicator has been defined as information about the characteristics and trends of ecosystem services, which may be used by policy-makers to understand the condition, trends and rate of change in ecosystem services (Layke, Mapendembe, Brown, Walpole, & Winn, 2012; Maes et al., 2016). Based on the adopted framework, two types of ecosystem service indicators have been used: indicators that communicate the *capacity* of an ecosystem to provide a service and those that communicate the *flow*, or actual provision, of an ecosystem service.

The case-study assessed the role of various terrestrial ecosystems in the delivery of multiple ecosystem services. Results demonstrate that semi-natural and agroecosystems make the backbone of the green infrastructure network in the Maltese Islands, and that these are important for the delivery of various ecosystem services leading to improved human well-being in Malta. In contrast, predominantly urban areas were characterised with a low capacity of ecosystems to provide services resulting in societal benefits affecting human well-being, and indicating that ecosystem service delivery in the landscapes of Malta is determined by land use intensity (Balzan, Caruana, et al., 2018).

The availability of green infrastructure and the contribution of this to the delivery of ecosystem services in each locality in the Maltese Islands was analysed by Balzan (2017). In this work, the area of green infrastructure in each local council was calculated from a generated land use land cover (LULC) map. Given that green infrastructure is considered as being a network of natural and semi-natural areas that provides a wide range of ecosystem services (EC, 2013), the cover of ecosystems contributing to the delivery of multiple ES was summed up for each locality. Ecosystem service capacity was strongly associated with the availability of green infrastructure, whilst urban areas associated with higher population densities had the lowest green infrastructure cover and ecosystem services capacity. In contrast, higher ecosystem service flow rates were recorded in ecosystems located in urban environments, which may be associated with higher ecosystem service demands in cities. Balzan et al. (2018a) recorded the highest $NO_2$ removal flux in woodland areas located in urban environments. Similarly, in a recent work assessing and mapping recreational ecosystem services through the use of georeferenced data from the GPS-outdoor game Geocaching[ii], the highest cache density was recorded in woodland and the urban green and leisure area category. The latter was also the most likely to be considered as a favourite point by those participating in this recreational game (geocachers). The results also indicate that the wider landscape impacts on ecosystem service flows, and the number of favourite points was positively associated with areas of high landscape value (Balzan & Debono, 2018).

## DEVELOPING EVIDENCE-BASED NATURE-BASED SOLUTIONS INTERVENTIONS

The described results demonstrate that the identified ecosystems contribute to the delivery of multiple ecosystem services within the study area. Other studies have similarly recorded positive associations between green infrastructure and ecosystem service delivery, for example, through improved food security (Dennis & James, 2016), the removal of air pollution and the provision of space for recreation (Baró et al., 2016), local climate regulation (Zardo, Geneletti, Pérez-Soba, & Van Eupen, 2017) and improved mental health (Alcock, White, Wheeler, Fleming, & Depledge, 2014; van den Berg et al., 2015). These results provide evidence that land use planning that implements and promotes the use of nature-based solutions whilst developing green infrastructure can significantly contribute to maintain biodiversity within landscapes whilst generating ecosystem services leading to benefits to human well-being.

Nature-based solutions are implemented to address societal challenges, and therefore selected methods and approaches will depend on local needs (Raymond et al., 2017). Assessments presented here provide a first evidence of the distribution of green infrastructure, and the benefits arising from the delivery of various ecosystem services. It may be complemented by the development of indicators for the assessment of ecosystem service demand, which can be compared with ecosystem service flow indicators to identify whether the demand for ecosystem services is satisfied (Baró et al., 2016). This would provide key information about the unsatisfied demand for ecosystem services leading to identified local challenges, for example as a consequence of high air pollution levels, surface water runoff generation due to soil sealing or limited availability of green spaces for the recreation of residents. Nature-based solutions are implemented to provide solutions to such challenges and therefore benefit from ecosystem service assessment and maps which build the scientific evidence about the needs for the implementation of nature-based solutions. In addition, challenges requiring nature-based solutions interventions impact on socio-ecological systems, are therefore often complex and multidimensional and, would therefore benefit from transdisciplinary and participatory approaches which offer informed societal choices. The alternative is the uninformed implementation of nature-based solutions without measures of their effectiveness in addressing societal challenges. The identification of nature-based solutions is normally based on an assessment of its potential to lead to environmental, social and economic benefits, through ecosystem services delivery. Consequently, indicators are chosen for the monitoring of the long-term impact of nature-based solutions at adequate geographical and ecological scales (Cohen-Shacham, Walters, Janzen, & Maginnis, 2016). This is an opportunity for practitioners, landscape and urban planners, and decision-makers to promote the development of the knowledge base about the links between green infrastructure and human well-being, and to guide ongoing and planned afforestation and restoration initiatives and the inclusion of other nature-based solutions, such as green walls and roofs, sustainable urban drainage systems and green spaces, in urban areas for more effective implementation.

CONCLUSION

This paper has provided an overview of the recent work mapping and assessing ecosystem services in Malta and the implications for decision-making for the implementation nature-based solutions to address societal challenges. Gradients in green infrastructure availability and in the delivery of ecosystem services have been identified. The mapping and assessment of ecosystem services provide information about the links and balances between the capacity and flow of ecosystem services and the arising benefits to human well-being. These play an important role in the development of the evidence basis for the identification of nature-based solutions leading to societal, environmental and economic impacts and which address societal challenges.

TABLES

*Table 1 – Defining the key components of an ecosystem services conceptual framework (adapted from Potschin-Young et al., 2018).*

| Term | Definition |
|---|---|
| Beneficiaries | A person or group whose well-being is changed in a positive way by an ecosystem service conservation. |
| Driver | Any natural or human-induced factor that indirectly causes a change in an ecosystem. |
| Ecosystem condition | The physical, chemical and biological condition or quality of an ecosystem at a particular point in time. |
| Ecosystem function | The subset of the interactions between biophysical structures, biodiversity and ecosystem processes that underpin the capacity of an ecosystem to provide ecosystem services. |
| Ecosystem service | The direct and indirect contributions of ecosystems to human well-being. |
| Ecosystem service capacity | The ability of an ecosystem to generate a specific ecosystem service in a sustainable way. |
| Ecosystem service demand | The expression of the beneficiaries' preferences for specific ecosystem service attributes, such as biophysical characteristics, location and timing of availability, and associated opportunity costs of use. |
| Ecosystem service flow | The actual production or use of the ecosystem service in a specific area and time |
| Ecosystem structure | A static characteristic of an ecosystem that is measured as a stock or volume of material or energy, or the composition and distribution of biophysical elements. |
| Governance | The process of formulating decisions and guiding the behaviour of humans, groups and organisations in formally, often hierarchically organised decision-making systems or in networks that cross decision-making levels & sector boundaries. |
| Green infrastructure | A strategically planned network of natural and semi-natural areas with other environmental features designed & managed to deliver a wide range of ecosystem services. It incorporates green or blue spaces and other physical features in terrestrial, coastal and marine areas. On land, green infrastructure is present in rural and urban settings. |
| Nature-based solutions | Living solutions inspired by, continuously supported by and using nature, which are designed to address various societal challenges in a resource-efficient and adaptable manner and to provide simultaneously economic, social, and environmental benefits. |
| Pressure | Human induced process that directly alter the condition of ecosystems. |

*Table 2 - Indicators used for the assessment and mapping of ecosystem services and based on previous contributions* (source: Balzan, 2018)

| Ecosystem Service (CICES 4.3) | Indicator | Capacity/Flow |
|---|---|---|
| **Cultivated crops** | Downscaled crop production (ton/Km$^2$) | Capacity/Flow |
| **Reared animals and their outputs** | Beekeepers' Habitat Preference (Frequency of responses) | Capacity |
| **Reared animals and their outputs** | Number of hives/Km$^2$ | Flow |
| **Materials from plants, algae and animals for agricultural use** | Rainfed agricultural land (Fodder production potential) | Capacity |
| **Materials from plants, algae and animals for agricultural use** | Livestock (number of Cattle, Sheep, Goats)/Km$^2$ | Flow |
| **Pollination and seed dispersal** | Pollinator visitation probability | Capacity |
| **Pollination and seed dispersal** | Crop pollinator dependency | Flow |
| **Dilution by atmosphere, freshwater and marine ecosystems** | NO$_2$ deposition velocity (mm/s) | Capacity |
| **Dilution by atmosphere, freshwater and marine ecosystems** | NO$_2$ removal flux (ton/ha/year) | Flow |
| **Physical use of land-/seascapes in different environmental settings** | Number of habitats of community importance | Capacity |
| **Physical use of land-/seascapes in different environmental settings** | Visitation to sites and urban green areas for recreational activities | Flow |
| **Physical use of land-/seascapes in different environmental settings** | Geocaching point location | Capacity |
| **Physical use of land-/seascapes in different environmental settings** | Number of geocache quests/favourites | Flow |
| **Aesthetic** | Preference Assessment with locals (Frequency of responses) | Flow |

FIGURES

*Figure 1: Conceptual framework for the assessment and mapping of ecosystem services to develop the evidence base for nature-based solutions interventions.*

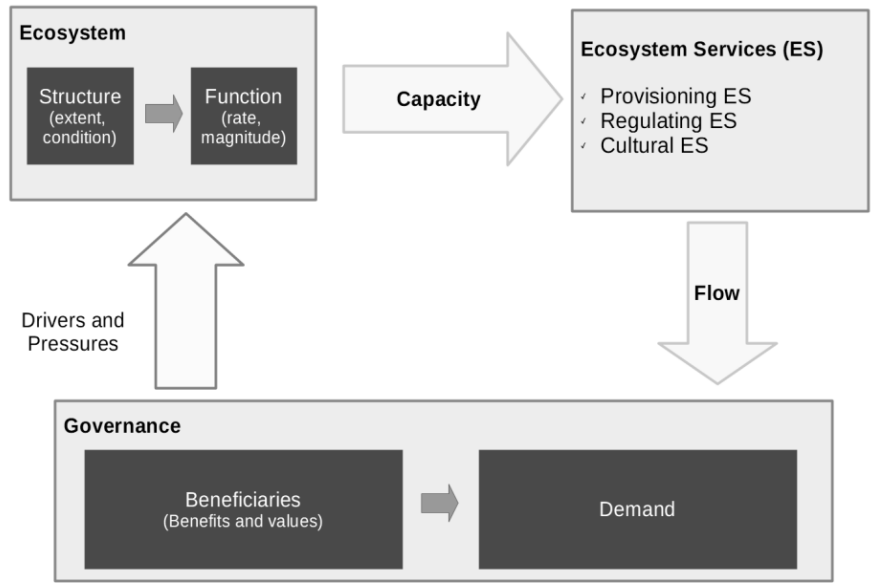

Adapted from: Balzan et al., (2018a)

*Figure 2: (a) Assessing the relationship between green infrastructure cover (GI) in each local council and average ES capacity and (b) population density.*

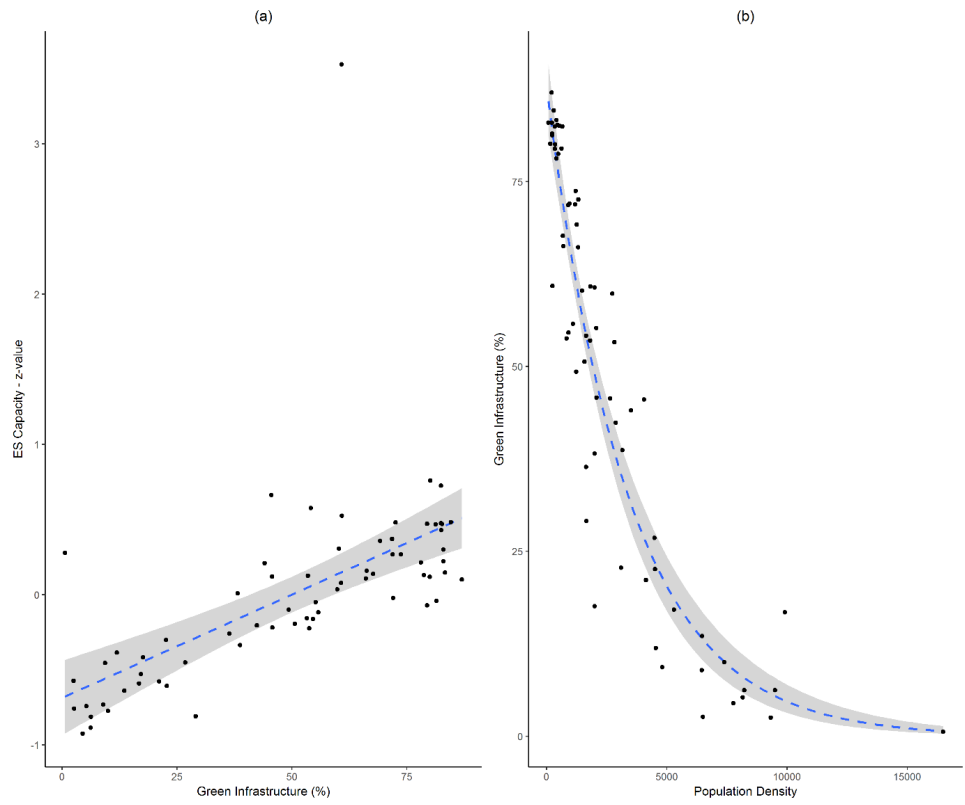

Source: Balzan (2017)

ACKNOWLEDGEMENTS

The author has received funding from the European Union's Horizon 2020 project ESMERALDA under grant agreement No 642007 and from the European Union Horizon 2020 project ReNature under grant agreement No 809988.

AFFILIATIONS

*Mario V Balzan*
*Institute of Applied Sciences,*
*Malta College of Arts, Science and Technology (MCAST)*
*Corradino Hill, Paola, PLA9032*
*Email: mario.balzan@mcast.edu.mt*

BIO

Mario V Balzan PhD, Senior Lecturer II at the Institute of Applied Sciences, Malta College of Arts, Science and Technology (MCAST). He has read for Ph.D. in agro-environmental and biodiversity science and a Master's degree in natural resources management. Dr Balzan has a multidisciplinary background in environmental and applied ecological sciences and has for several implemented research about ecosystem services and nature-based solutions across rural and urban landscapes. He is the coordinator of the Horizon 2020 project ReNature and the Principal Investigator of the MCAST Applied Environmental Sciences Research Group (AESReG), where he coordinates the research activities on biodiversity, ecosystem services and nature-based solutions. Research carried out by the group addresses sustainability issues that would benefit from multidisciplinary approaches to environmental management and policy-making. Past research has investigated biodiversity-ecosystem function relations, the assessment and mapping of ecosystem services, and the provision of nature-based solutions for maintaining ecosystem services in cultural landscapes. This research has been supported through projects that have received funding from national and international funding agencies.

NOTES

[i] http://www.esmeralda-project.eu
[ii] https://www.geocaching.com/play